\documentclass[letter]{aa}     % for the letters
\usepackage{graphicx}
\usepackage{txfonts}
\usepackage{lipsum}
\usepackage{subcaption}         % necessary for continued figures, example in section 3
\usepackage{lscape}             % to rotate a single page table, example in appendix.
\usepackage{placeins}           % useful with \FloatBarrier, to keep 
\begin{document}

   \title{Dust-driven streaming instability and magnetic field amplification downstream of supernova remnant shocks}

   %\subtitle{Subtitle}

%%%%%%%%%%%%%%%%%%%%%%%%%%%%%%%%%%%%%%%%
% Please do not include ORCIDs next to author names.
% Only ORCIDs authenticated by individual authors in EDP Sciences editorial system will be taken into account.
% ORCIDs included here will be removed.
%%%%%%%%%%%%%%%%%%%%%%%%%%%%%%%%%%%%%%%%

   \author{S. Gabici\inst{1}%\fnmsep\thanks{Shows the usage of elements in the author field}
        \and J.~C. Raymond\inst{2}
        \and A. Ciardi\inst{3}
        \and P. Cristofari\inst{3}
        \and S. Recchia\inst{4,5}
        \and V. Tatischeff\inst{6}
        }

   \institute{Universit\'e de Paris Cit\'e, CNRS, Astroparticule et Cosmologie,  F-75006 Paris, France
             %\email{gabici@apc.in2p3.fr}
             %\thanks{Shows the usage of elements in the author field}
            \and Center for Astrophysics, 60 Garden St., Cambridge, MA 02176, USA
            %\and Sorbonne Universit\'e, Observatoire Paris, Universit\'e PSL, CNRS, Laboratoire d’\'Etude de l’Univers et des Ph\'enom\`enes Extr\^emes, LUX, F-75005 Paris, France
            \and LUX, Observatoire de Paris, Universit\'e PSL, Sorbonne Universit\'e, CNRS, 92190 Meudon, France
            \and Istituto Nazionale di Astrofisica – Osservatorio Astrofisico di Arcetri, Largo Enrico Fermi 5, 50125, Firenze, Italy
            \and (IFJ-PAN)
Institute of Nuclear Physics Polish Academy of Sciences, PL-31342 Krakow, Poland
            \and Universit\'e Paris-Saclay, CNRS/IN2P3, IJCLab, 91405 Orsay, France}

   \date{Received September 30, 20XX}

% \abstract{}{}{}{}{}
% 5 {} token are mandatory
 
  \abstract
   {The acceleration of cosmic rays up to PeV energies at supernova remnant shocks requires an amplification of the ambient magnetic field.
The amplification mechanism must operate upstream of the shock, to prevent the escape of particles from the system.
Observational evidence of field amplification has been indeed obtained by means of X-ray observations.
However, such observations constrain the magnetic field strength downstream of the shock only.}
   {Here we describe a mechanism for magnetic field amplification that operates downstream of the shock. It is based on a plasma instability triggered by the drift of charged interstellar dust grains overtaken by the shock.} 
   {We compute the growth rate of the instability, we estimate the level of magnetic field amplification expected downstream of supernova remnant shocks, and we compare our results with observations.}
   {In some cases (most notably Cas~A) this mechanism might explain the presence of the X-ray filaments observed at supernova remnant shocks, without requiring any amplification of the magnetic field upstream of the shock and therefore no acceleration of CRs to ultra-high energies.
%The large magnetic field strength measured at Cassiopea~A, and possibly other objects, could be explained in this way.
}
   {}


   \maketitle

\nolinenumbers 

\section{Introduction}

Galactic cosmic rays (CRs) 
are believed to be accelerated 
at supernova remnant (SNR) shocks~\citep{gabici2019}.
However, this scenario is challenged by a number of observations. 
In particular, it is not clear whether SNR shocks can accelerate protons up to the PeV energy domain, as required by 
near-Earth observations of CRs.
This is why the search for proton PeVatrons in the Galaxy is a major goal of gamma-ray astronomy \citep{cao2021}. 
Despite tremendous progress, this remains an open issue. 

The amplification of the magnetic field (MF) to values exceeding the interstellar one is a necessary condition for acceleration to PeV energies at SNRs~\citep{hillas2005}.  
The synchrotron X-ray filaments observed at some SNR shocks demonstrate that the MF there is indeed largely amplified~\citep{helder2012}. 
The amplification may be due to plasma instabilities triggered by the streaming of CRs accelerated at the shock~\citep{marcowith2021}.
In this scenario, the amplification takes place upstream of the shock, it improves the confinement of CRs within the acceleration region, and boosts their energy to large values.

Here we show that interstellar dust grains overtaken by a shock trigger a plasma instability that may result in a large amplification of the MF.
In this case, the amplification takes place downstream of the shock and therefore does not improve the confinement of CRs at SNRs, nor increases their maximum energy.
After comparing our predictions with observations, we conclude that this scenario may be a viable alternative to explain the large values of the MF strength observed in some SNRs.

\section{Charged dust grains overtaken by shocks}
\label{sec:grains}

Consider dust grains of mass $M_g = A_g m_p$ and charge $Q_g = Z_g e$, where $m_p$ and $e$ are the proton mass and charge, 
and $A_g$ and $Z_g$ the grain's analog of atomic mass and number, respectively.
For simplicity, we assume that grains are spheres of radius $a$.

Interstellar dust grains acquire a positive charge due to a number of processes such as photoelectric effect, impact of atoms or ions on the grain's surface, etc.~\citep{ellison1997}.
As a result, grains acquire an electric potential $\phi = Q_g/a$ of the order of 10-100~V or, equivalently, a charge:
\begin{equation}
\label{eq:Zg}
Z_g = \frac{a \phi}{e} \sim 6.9 \times 10^2 \left( \frac{a}{10^{-5} ~ {\rm cm}} \right) \left( \frac{\phi}{10~{\rm V}} \right) ~ .
\end{equation}
The grain atomic number is:
\begin{equation}
A_g = \frac{4 \pi}{3} a^3 \frac{\varrho_g}{m_p} \sim 7.5 \times 10^9 \left( \frac{\varrho_g}{3~{\rm g/cm^{3}}} \right) \left( \frac{a}{10^{-5}~{\rm cm}} \right)^3 
\end{equation}
where $\varrho_g$ is the density of grain material \citep{cristofari2025}.

Consider now grains swept up by a shock of velocity $u_s$.
Initially, grains are not affected by the shock passage, as they are characterised by rigidities much larger than those of thermal particles. 
Therefore, once downstream they move with a velocity $u_g = (3/4) u_s$ with respect to the shocked gas.
In the rest frame of the downstream fluid the momentum of dust grains is $p_g =A_g m_p u_g$, corresponding to PetaVolt rigidities
($R = p_g c/Z_g e$)
and sub-parsec scale Larmor radii ($r_L = R/B_0)$
\begin{equation}
\label{eq:rigidity}
\frac{R}{\rm PV}
\sim 0.13 \left( \frac{u_s}{5 \times 10^8 {\rm cm/s}}\right)  \left( \frac{\varrho_g}{3~{\rm g/cm^{3}}} \right) \left( \frac{a}{10^{-5}~{\rm cm}} \right)^2 \left( \frac{10~{\rm V}}{\phi} \right)
\end{equation}
\begin{equation}
\label{eq:larmor}
\frac{r_L}{\rm pc}
\sim 0.05  \left( \frac{u_s}{5 ~ 10^8 {\rm cm/s}}\right) \left( \frac{\varrho_g}{3~{\rm g/cm^{3}}} \right) \left( \frac{a}{10^{-5} {\rm cm}} \right)^2 
\left( \frac{10~{\rm V}}{\phi} \right) \left( \frac{3~\mu {\rm G}}{B_0} \right) \nonumber
\end{equation}
where $B_0$ is the MF strength downstream of the shock.

Note that, especially for young SNRs characterised by large shock speeds and small sizes, the Larmor radius of grains can be of the same order of the thickness of the SNR shell, which is a small fraction of the SNR radius.
When that happens, grains are unmagnetised and cross the region suffering very little deflection.
The current associated to the streaming of charged dust grains excites plasma instabilities similar to those triggered by CR streaming (\citealt{hopkins2018}).
Given the similarities in the behavior of CR nuclei and charged grains, it  is natural to wonder what contribution the latter provide to the amplification of MFs at shocks.
Here we argue that, in some cases, dust grains might play a major role in amplifying the MF downstream of the shock, and not upstream, as CRs are believed to do.

Further downstream of the shock, grains will be eventually isotropised by the MF and, due to their large rigidities, will behave as CR nuclei. 
Some of them may be able to recross the shock and start to be accelerated via first order Fermi mechanism.
This has been described in, e.g.,
\citet{ellison1997,cristofari2025}, and will not be further discussed here.

\section{Dust-driven streaming instability}
\label{sec:instability}

The growth rate of the dust-driven instability is estimated here by means of an analogy with CRs. 
CR streaming instability operates at shocks in two flavors:  resonant 
and non-resonant 
mode~\citep{marcowith2021}.
For simplicity, we limit our analysis to parallel shocks, where the direction of the ordered component of the MF in the upstream plasma is parallel to the shock normal.
Both streaming instabilities are driven by a $\bf{j} \times \bf{B}$ force, where $\bf{j}$ is a current resulting from the streaming of charged particles, and $\bf{B}$ is the ambient MF.
When a small perturbation ${\bf{B}}_1$ is added perpendicular to the ordered background field ${\bf{B}}_0$, the current will be also perturbed as ${\bf j} = {\bf j_0} + {\bf j_1}$, where $j_1 \ll j_0$ and ${\bf j}_0 \parallel {\bf B}_0$.
The resonant instability is driven by the ${\bf j}_1 \times {\bf B}_0$ force \citep{kulsrud1969}, while the non-resonant by the ${\bf j}_0 \times {\bf B}_1$ one \citep{bell2013}.
The non-resonant instability grows faster, and therefore we discuss it further.

Assume that all dust grains are identical and characterised by the same size, mass, and charge, and that their density in the interstellar medium is $n_g$.
A generalization to the case of a broad distribution of sizes (and therefore masses and charges) is provided in Appendix~\ref{app:grains}. 
We anticipate that larger grains are more effective in triggering the instability.
As the size distribution of grains roughly scales as $n_g(a) \propto a^{-3.5}$ \citep{mathis1977} their integral mass distribution is proportional to $a \times n_g(a) \times A_g \propto a^{0.5}$ and is dominated by the largest grains.
Therefore, the grain mass density in the ambient medium $\varrho_{dust} = n_g A_g m_p$ is the parameter determining the effectiveness of the instability.

Dust grains crossing the shock surface from upstream generate a current in the downstream frame equal to:
\begin{equation}
\label{eq:jdust}
j_0  = n_g Q_g u_g = \left( \frac{3e}{4}\right) Z_g n_g u_s ~ .
\end{equation}
The current is parallel to the shock normal and directed towards downstream.
As dust grains are characterised by very large rigidities, their Larmor radius can be safely assumed to be much larger than the wavelength of perturbations of the MF.
Under these circumstances, $j_1 = 0$, and the equation of motion for the background thermal gas reads:
\begin{equation}
\varrho \frac{\partial {\bf u}_{\perp}}{\partial t} = - \frac{1}{c} ~{\bf j}_0 \times {\bf B}_{1}
\end{equation}
where ${\bf u}_{\perp}$ is the velocity of the plasma orthogonal to ${\bf B}_0$, $\varrho$ its (constant) density, and the minus sign on the right hand side indicates that the force is induced by a return current carried by the background plasma to balance ${\bf j}_0$.
When combined with the ideal MHD equation for the evolution of the MF:
\begin{equation}
\frac{\partial {\bf B}_1}{\partial t} = \nabla \times \left( {\bf u}_{\perp} \times {\bf B}_0 \right) 
\end{equation}
it gives the two equations:
\begin{equation}
\frac{\partial^2 B_x}{\partial t^2} = \frac{B_0 j_0}{\varrho c} \frac{\partial B_y}{\partial z}  ~~,~~
\frac{\partial^2 B_y}{\partial t^2} = -\frac{B_0 j_0}{\varrho c} \frac{\partial B_x}{\partial z}
\end{equation}
where $z$ ($x, y$) indicate the directions parallel (perpendicular) to ${\bf B}_0$.
These equations show that harmonic ($\propto e^{ikz}$) and right-hand circularly polarised modes are unstable and grow at a rate:
\begin{equation}
\label{eq:growth}
\gamma = \left( \frac{j_0 B_0 k}{\varrho c} \right)^{1/2} ~ .
\end{equation}

A distinct signature of the instability is that small scale (large $k$) perturbations grow faster.
The smallest possible scale at which the instability operates is determined by the equilibrium between the Lorentz force ($j_0 \times B_1/c$) which drives the instability, and the magnetic tension of field lines ($(B_0 \cdot \nabla) B_1/4 \pi$) which opposes to that.
This condition is satisfied for $k_{max} \sim (4 \pi / c) j_0/B_0$, which can be substituted in Eq.~\ref{eq:growth} to give the fastest growth rate:
\begin{equation}
\label{eq:gammamax}
\gamma_{max} \sim \frac{j_0}{c} \left( \frac{4 \pi}{\varrho} \right)^{1/2} ~ .
\end{equation}

If we now impose that the scale of the perturbation should be smaller than the grain Larmor radius (otherwise grains would gyrate around the perturbed field and the Lorentz force would average to zero) we obtain a condition on the current, $j_0 > B_0^2 c/4 \pi R$, which can be inverted to obtain the maximum possible value of the MF, $B_{sat}$, at which the instability saturates.
This can be written in a convenient form introducing the ambient (interstellar) dust-to-gas ratio $\chi = \varrho_{dust}/\varrho_{gas}$, giving:
\begin{equation}
\label{eq:handy}
\frac{B_{sat}^2}{8 \pi} \sim \frac{9}{32}~ \chi~  \varrho_{gas} u_s^2 ~ .
\end{equation}
The equation shows that the magnetic pressure downstream is a small fraction of the shock ram pressure, and that the proportionality constant is of the order of the dust-to-gas mass ratio.
Type Ia supernovae (SNe) explode at random places in the interstellar medium, so they would encounter gas with a typical dust-to-gas ratio $\chi \lesssim 0.01$\citep{jenkins2009}.  
The value of $\chi$ is more uncertain for core collapse SNe. SNe with red supergiant progenitors would encounter a dusty circumstellar medium until they grow beyond the size of the progenitor star wind, which could be on the order of $\approx$~1 pc \citep{vandyk2025}.
SNe from stripped progenitors, Type Ib/c, might or might not encounter dusty circumstellar material.
Normalising $\chi$ to the typical interstellar value we get:
\begin{equation}
B_{sat} \sim 2 \times 10^2 \left( \frac{\chi}{0.01} \right)^{1/2} \left( \frac{n_H}{\rm cm^{-3}} \right)^{1/2} \left( \frac{u_s}{5 \times 10^3  {\rm km/s}} \right) ~\mu \rm G
\end{equation}
where $n_{\rm H}$ is the hydrogen density in the interstellar medium (of mean atomic weight $\mu \sim 1.4$).
Remarkably, this is quite close to the MF strengths inferred from the observations of X-ray filaments at SNR shocks~\citep{helder2012}.

So far, we implicitly assumed that the background plasma is cold, which means that the Larmor radius of thermal particles $r_{L,i}$ ($i = p$ or $e$ for protons and electrons, respectively) are much smaller than $1/k_{max}$, or $r_{L,i} k_{max} \rightarrow 0$.
However, the plasma downstream of a SNR shock is heated to a large temperature $k_B T = (3/16) m_p u_s^2$, where $k_B$ is the Boltzmann constant, resulting in a large thermal velocity of protons $v_{T,p} = \sqrt{2 k_B T/m_p} = \sqrt{3/8} u_s$ and a correspondingly large Larmor radius.
Then one should rather consider a situation where $r_{L,p} k_{max}$ is small but not vanishing. 
The Larmor radius of thermal electrons can still be set equal to zero as $r_{L,e} \ll r_{L,p}$.
A finite Larmor radius means that protons in the plasma are less magnetised, and this results in a reduction of both the instability growth rate $\gamma_{max}$ and the wavenumber $k_{max}$ \citep[][]{reville2008,marret2021}.

It can be shown that thermal effects are important if the plasma Alfv\`en speed is \citep{zweibel2010}:
\begin{equation}
v_A < \left( Z_g \frac{n_g u_g}{n_p v_{T,p}} \right)^{1/3} v_{T,p}
\end{equation}
where $n_p$ is the downstream ambient proton density.
After some manipulations, and making use of Eq.~\ref{eq:handy}, this can be rewritten as a condition on the MF immediately downstream of the shock (i.e. before the dust-driven field amplification):
\begin{equation*}
B_d < \left( \frac{4}{3} \frac{Z_g}{A_g} \right)^{1/3} \frac{B_{sat}}{\chi^{1/6}}    
\sim
1 \left( \frac{A_g/Z_g}{10^7} \right)^{-1/3} \left( \frac{\chi}{0.01} \right)^{-1/6} \left( \frac{B_{sat}}{100~\mu{\rm G}} \right) \rm \mu G 
\end{equation*}
which shows that thermal effect are probably quite small. This is because the minimum possible value of $B_d$ can be estimated by assuming that no field amplification operates upstream. In that case $B_d$ will be equal to the upstream MF for a parallel shock and four times that for a perpendicular strong shock.
In both cases one expects values exceeding the microGauss level.
For this reason, in the following we neglect thermal effects.

Finally, the instability operates provided that dust grains are not decelerated due to friction with the background plasma \citep{draine1979}. 
\citet{ellison1997} showed that direct collisions between grains and plasma ions dominate the friction and that the corresponding momentum-loss time is: $\tau_{loss} \approx A_g/\mu a^2 n_p u_g$.
For typical parameters, this corresponds to stopping length $\tau_{loss} u_g$ of the order of few parsecs, which is orders of magnitude larger than the width of X-ray filaments ($\Delta l_X \approx 10^{17}$~cm, \citealt{helder2012}).
This shows that friction can be neglected and that the instability will lead to an amplification of the MF.

As the amplification mechanism operates downstream, a shift will exist between the position of the shock and the peak of the MF strength, and as a consequence also a shift between the shock position and the X-ray filament, or a broadening of X-ray filaments (depending on the exact profile of the MF downstream).
When the dust-driven instability is responsible for the amplification, the strength of the MF perturbation grows by a factor of $e$ in a time $\gamma_{max}^{-1}$.
The time needed to reach saturation is then $\tau_B = N_e/\gamma_{max}$, where $N_e$ represents the number of $e$-foldings of the MF.
Heuristic arguments suggest that $N_e \approx 5$ \citep[][]{schure2013}.
In a time $\tau_B$
the downstream fluid will move away from the shock at a speed $u_s/4$ and therefore the shift will be $\Delta l_B \approx \tau_B u_s/4$, or:
\begin{equation*}
\Delta l_B \approx 10^{17} \left( \frac{N_e}{5} \right) \left( \frac{10^{-12}\rm cm^{-3}}{n_g}\right) \left( \frac{n_H}{\rm cm^{-3}} \right)^{1/2} \left(\frac{10^{-5}\rm cm}{a} \right) \left( \frac{10~\rm V}{\phi} \right) ~ \rm cm.
\end{equation*}

The presence of a small (arcseconds) shift between the SNR shock and the X-ray emission might be detected by means of the observation of H$\alpha$ filaments that do trace the position of the shock \citep{heng2010}.
Unfortunately, cospatial H$\alpha$ and X-ray filaments are rare \citep{heng2010,knezevic2017,sankrit2008,sapienza2024}, and to our knowledge no combined analysis of their spatial profiles has been performed.

\section{Comparison with observations}

\begin{figure}
\centering
\includegraphics[width=0.8\columnwidth]{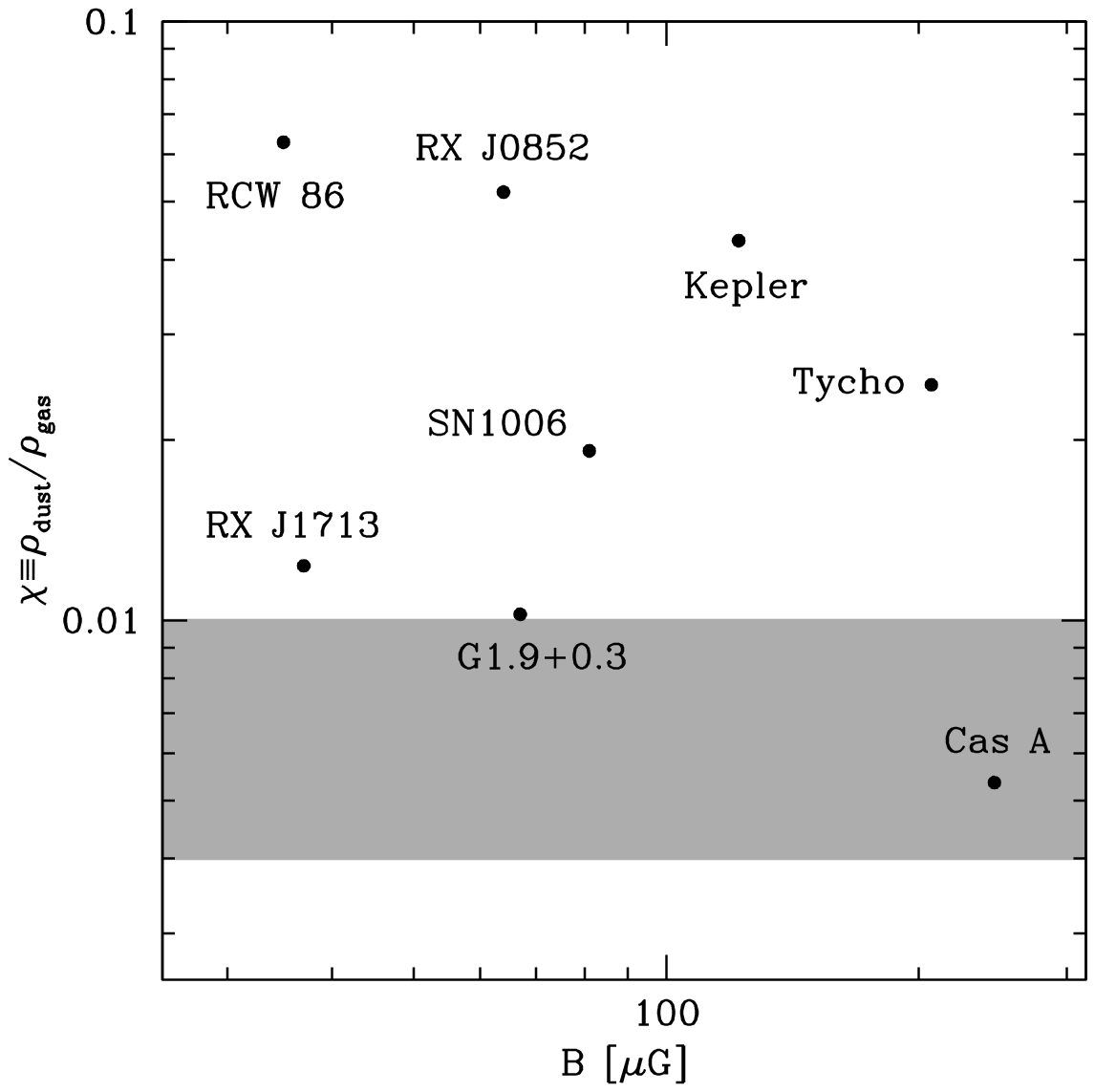}% Here is how to import EPS art
\caption{\label{fig:helder} Dust-to-gas mass ratio ($\chi$) needed to explain the observed values of the MF strength downstream of some SNRs. The shaded region indicates typical interstellar values for $\chi$.}
\end{figure}

In order to compare our predictions with data we notice that, with the exception of $\chi$, all quantities appearing in Eq.~\ref{eq:handy} are measurable.
Therefore, we made use of the compilation of data for 8 young Galactic SNRs from \citet{helder2012}, and derived the value of the dust-to-gas mass ratio one would need to explain the observed MFs in terms of the dust-driven instability discussed here.
The values of $\chi$ obtained in this way are plotted in Fig.~\ref{fig:helder}, where the shaded region indicates the range of typical interstellar values \citep[][]{jenkins2009}, which are appropriate at least for SNRs with a type Ia progenitor.

Fig.~\ref{fig:helder} shows that, in order to explain the observed MF strengths, $\chi$ should be in the range $\approx 0.5-6$\%.
Cas~A is the only object for which $\chi$ is smaller than 1\%, for G1.9+0.3 $\chi \sim 0.01$, and for all the other objects larger values are required.
G1.9+0.3 and Cas~A are the two youngest SNRs in our sample, the former is most likely the result of a type Ia SN \citep{chakraborti2016}, while the latter a type IIb \citep{krause2008}.
As Cas~A is the SNR characterised by the largest value of the MF and also by the smallest (and very plausible) required value of the dust-to-gas ratio, we discuss it in more detail.

Cas~A exploded about 350 years ago 
and its forward shock is sweeping up dusty gas over a substantial part of its area \citep{delooze2024}.  However, not all
of the circumstellar gas is dusty.  
\citet{koo2020} studied at high spectral resolution a circumstellar clump which has not yet been processed by the SNR shock
and found that
iron was undepleted in the photoionization precursor, suggesting an absence
of dust.
Therefore, even though further studies are needed to determine the circumstellar value of the dust-to-gas mass ratio, a large value of $\chi$ is not implausible, at least in a sizeable fraction of the upstream material.

Competing mechanisms for MF amplification must be discussed.
In particular, CRs escaping from Cas~A can amplify the MF upstream of the shock \citep{bell2013}.
In order to assess the importance of such mechanism we notice that the gamma-ray emission observed from Cas~A is strongly suppressed above a photon energy of $E_{\gamma}^{max} \approx 1$~TeV \citep{cao2025}.
If the emission is the result of the decay of neutral pions produced in inelastic interactions of CR protons with ambient matter, then the spectrum of accelerated protons cuts off at  $E_{max} \approx 10$~TeV.
This very fact implies that dust grains of larger (sub-PV, see Eq.~\ref{eq:rigidity}) rigidities will not be scattered effectively by MF turbulence.

The maximum energy of protons accelerated via diffusive shock acceleration at a SNR can be estimated by equating their diffusion length upstream of the shock $D(E)/u_s$ to a small fraction of the shock radius $\eta R_s$.
Here, $D(E) = (1/3) r_L c$ is the Bohm CR spatial diffusion coefficient and the geometric factor $\eta$ is small and often assumed to be $\approx 0.05$ \citep[][]{ptuzir2005}.
This gives a value for the upstream MF equal to:
\begin{equation}
\label{eq:Bup}
\frac{B_{up}}{\mu {\rm G}} \approx 1.4 ~\left( \frac{\eta}{0.05} \right)^{-1} \left( \frac{R_s}{3~{\rm pc}} \right)^{-1} \left( \frac{u_s}{5000~{\rm km/s}} \right)^{-1} \left( \frac{E_{max}}{10~{\rm TeV}} \right) 
\end{equation}
where we normalized physical quantities to values appropriate for Cas~A \citep[][]{helder2012}.
This value is quite small and indicates that mechanisms of MF amplification operating upstream of the shock (not necessarily CR-driven, see e.g. \citealt{beresnyak2009}) are not very effective.
Compression at the shock may enhance the downstream MF above the value indicated in Eq.~\ref{eq:Bup}.
However, the large scale MF in Cas~A is predominantly radial \citep{mercuri2025}, thereby limiting the impact of shock compression.
This suggests that the MF amplification may well operate downstream of the shock.

Several MF amplification mechanisms may operate downstream of shocks.
Vorticity is produced downstream of a warped shock interacting with a medium characterized by large scale density fluctuations.
Such vorticity can in turn distort, stretch, and eventually amplify the MF \citep{giajok2007}.
Amplification could also result from the advection of magnetic turbulence through the shock \citep{zank2021}.
Note that these mechanisms do not require the presence of CRs, but may still compete with the dust-driven mechanism proposed in this paper.

We conclude discussing another interesting object: the type IIb SN~1993J, for which estimates of the MF have been obtained based on radio observations \citep{fransson1998}, 
implying
a scaling in time of the downstream MF of the form $B_d = B_* (t/{\rm days})^{-1}$, with $B_* \approx 180$~G \citep{tatischeff2009}. 
Note that this would be the same scaling predicted by Eq.~\ref{eq:handy} for a shock moving at a speed $u_s \propto R_s/t$ 
across the progenitor stellar wind of density profile $\varrho_{gas} \propto R^{-2}$, provided that the dust-to-gas ratio $\chi$ is uniform in space.
An order of magnitude estimate of $\chi$ can be obtained by assuming that the progenitor stellar wind is characterised by a mass loss rate of $\dot{M} \approx 5 \times 10^{-5} M_{\odot}$/yr and a speed $u_w \approx 10$~km/s. 
Recalling that $\varrho_{gas} = \dot{M}/4 \pi u_w R^2$ and making use of Eq.~\ref{eq:handy} we get $\chi \approx 0.1$, which is quite large.
Therefore, unless the progenitor wind is very dusty, other amplification mechanisms have to be invoked in this case.

\section{Discussion and conclusion}

We have shown that dust-driven streaming instability can operate downstream of SNR shocks.
In some cases the resulting MF amplification might explain the presence of the narrow X-ray filaments observed at SNR shocks. 
In this scenario, the MF upstream of the shock is not amplified, CRs are confined less effectively, and this prevents their acceleration to PeV energies.
Cas~A fits particularly well within this framework, being characterised by a strong MF downstream of the shock ($\sim 250 ~\mu$G) and a low value of the maximum energy of accelerated CRs ($\sim 10$~TeV).
The instability might also explain the large values of the MF observed in other SNRs, most notably G1.9+0.3.

The main point raised in this paper is the fact that, under some circumstances, the observation of strong MFs at SNR shocks does not necessarily imply that favorable conditions for acceleration of CRs to ultra-high energies are present.
However, for systems different than SNRs, such as stellar wind termination shocks, the situation can be different.
In this case, the wind blown by a star is decelerated at a slowly expanding termination shock \citep{morlino2021,gabici2024}. 
Therefore, the downstream region is outside of the spherical shock (and not inside, as in SNRs).
For dusty stellar winds, the instability discussed in this paper might indeed improve the confinement within the acceleration region (i.e. the wind termination shock), and possibly boost acceleration of particles to larger energies.
It would be interesting to extend the present study to these systems.

Finally, as the MF amplification takes place downstream, it might facilitate the reflection of particles back towards the shock upstream and play a role in the complex process of injection into the acceleration mechanism \citep{caprioli2014}.

We plan to go beyond the analytic approach presented here, and perform numerical simulations (hybrid particle-in-cell codes, \citealt{aunai2024}) to better characterise the instability and its effects on SNR shock environments.

\begin{acknowledgements}
We thank F. Calura, J. Duprat, M. Lemoine, M. Miceli, G. Morlino, and B. Schroer for useful discussions.
\end{acknowledgements}

\bibliographystyle{aa} 
\bibliography{biblio}

\begin{appendix}

\section{Size distribution of dust grains}
\label{app:grains}

Here we generalize Eq.~\ref{eq:handy} to the case of non identical grains. Two parameterizations are widely used in the literature to describe the size distribution of grains: a scale free power-law distribution \citep{mathis1977}, and a flatter, non-scale-free, parameterization \citep{weingartner2001}.
The latter requires to specify more free parameters but is more accurate as it reproduces the extinction of stellar light in the Milky Way. 
The former is most likely oversimplified but still useful to perform order of magnitude estimates.

\citet{mathis1977} assumed that  the size distribution of grains is $n(a) \propto a^{-3.5}$ in the range $a_{min} \equiv 5 \times 10^{-7}~{\rm cm} < a <2.5 \times 10^{-5} {\rm cm} \equiv a_{max}$.
One can see from Eq.~\ref{eq:rigidity} that the smallest grains are characterized by small Larmor radii, and therefore do not contribute to the current as they are easily isotropized.
However, such small grains carry a little fraction of the total mass of interstellar dust. 
We therefore introduce a minimum size $a_{j}$ of grains which are substantially contributing to the current. When this is done the mass density of interstellar mas dust-to-gas ratio can be written as:
\begin{equation}
\chi \equiv \frac{\varrho_{dust}}{\varrho_{gas}} = \frac{1}{\varrho_{gas}} \int_{a_{min}}^{a_{max}} {\rm d} a ~ n(a) ~ \frac{4 \pi}{3} a^3 \varrho_g \propto a_{max}^{0.5}
\end{equation}
while the current carried by grains reads (Eq.~\ref{eq:jdust}):
\begin{equation}
j_0 = \left( \frac{3 e}{4} \right) u_s \int_{a_{j}}^{a_{max}} {\rm d} a ~ Z_g(a) ~ n(a) \propto a_{j}^{-1.5}
\end{equation}
where the proportionalities are to be considered as approximate and are reported only to illustrate that the largest grains carry most of the mass while the smallest ones (among those which are not isotropized) carry most of the current (exact expressions will be used below).
This statement holds even more strongly if one adopts the grain size distribution by \citet{weingartner2001}, which is flatter.

At this point, we can follow the same rationale as in Sec.~\ref{sec:instability} and assume that the instability is quenched when the Larmor radius of dust grains in the amplified MF becomes of the same size of the perturbation, and therefore grains are isotropized.
A difference with respect to the case of identical grains is that when this condition is satisfied for grains of size $a_j$, larger grains will still contribute to the current.
We therefore assume that the instability stops when the current drops by a factor of $e$.
Under these assumptions Eq.~\ref{eq:handy} can be rewritten as:
\begin{equation}
\label{eq:corrected}
\frac{B_{sat}^2}{8 \pi} \sim \frac{9}{32}~ f(a_j) ~\chi~  \varrho_{gas} u_s^2 
\end{equation}
where $f(a_j)$ is a correction factor that accounts for the broad distribution of dust grain sizes and reads:
\begin{equation}
f(a_j) = \frac{1}{3} a_j^2 \frac{a_j^{-3/2}-a_{max}^{-3/2}}{a_{max}^{1/2}-a_{min}^{1/2}} \left[ \left( 1 - \frac{1}{e}\right) \left( \frac{a_j}{a_{max}} \right)^{3/2} + \frac{1}{e} \right]^{-4/3}
\end{equation}
which reaches a maximum value $f \sim 0.50$ at $a_{j,max} \sim 0.57 \times 10^{-5}$~cm.
The maximum of the function is broad, and for $a_{j,ref} = 10^{-5}$~cm (the reference grain size adopted in the paper) one gets $f \sim 0.43$.
If we use Eq.~\ref{eq:corrected} instead of Eq.~\ref{eq:handy} to derive the value of $\chi$ needed to explain the MF strength measured in Cas~A, we get $\chi \sim 1.1 - 1.2$\% for $a_{j,max} < a < a_{j,ref}$.

Smaller values of $\chi$ are expected if the flatter and more accurate grain size distribution from \citet{weingartner2001} is adopted.
This is because the resulting mass distribution is even more peaked towards large grains.
Even though this distribution depends on several free parameters, for large values of the grain size it roughly behaves as a power-law plus an exponential cutoff, with an index of the power law which is in general smaller than 3.5, and with a maximum (cutoff) size that might even be larger than the $a_{max}$ fixed by \citet{mathis1977}.
Both this things would reduce the value of $\chi$ required to explain the large MF observed in Cas~A.

As we are interested in large grains only (small grains carry a small fraction of the mass and most likely do not contribute to the current as they are easily isotropized), to mimic the effects of a more realistic size distribution of grains we repeat the calculation above for a flatter size distribution $n(a) \propto a^{-3}$, keeping the value of $a_{max}$ unchanged (assuming a larger value would lower even more the value of $\chi$).
When this assumption is done, one gets $\chi \propto a_{max}$ and $j_0 \propto a_{j}^{-1}$. Repeating the procedure above for Cas~A one gets $a_{j,max} = 0.67 \times 10^{-5}$~cm and values of the dust-to-gas mass ratio $\chi = 0.77-0.84$\% for $a_{j,max} < a < a_{j,ref}$.

All the values of $\chi$ obtained above are plausible, and within less than a factor of $\sim 2$ from what obtained by using Eq.~\ref{eq:handy}, which is therefore quite accurate.

\end{appendix}

\end{document}